\documentclass[apl,superscriptaddress,reprint,amssymb,aps,floatfix,showkeys]{revtex4-1}
\usepackage{amsmath}%
\usepackage{amsfonts}%
\usepackage{amssymb}%
\usepackage{graphicx}
\usepackage{color}
\usepackage{tabularx}
\usepackage{braket}

\begin{document}
\title{Observation of the Fundamental Nyquist Noise Limit in an Ultra-High $Q$-Factor Cryogenic Bulk Acoustic Wave Cavity}

\author{Maxim Goryachev}
\email{maxim.goryachev@uwa.edu.au}
\affiliation{ARC Centre of Excellence for Engineered Quantum Systems, University of Western Australia, 35 Stirling Highway, Crawley WA 6009, Australia}

\author{Eugene N. Ivanov}
\affiliation{ARC Centre of Excellence for Engineered Quantum Systems, University of Western Australia, 35 Stirling Highway, Crawley WA 6009, Australia}

\author{Frank van Kann}
\affiliation{School of Physics, University of Western Australia, 35 Stirling Highway, Crawley WA 6009, Australia}

\author{Serge Galliou}
\affiliation{Department of Time and Frequency, FEMTO-ST Institute, ENSMM, 26 Chemin de l'\'{E}pitaphe, 25000, Besan\c{c}on, France}

\author{Michael E. Tobar}
\affiliation{ARC Centre of Excellence for Engineered Quantum Systems, University of Western Australia, 35 Stirling Highway, Crawley WA 6009, Australia}

\begin{abstract}

Thermal Nyquist noise fluctuations of high-$Q$ Bulk Acoustic Wave (BAW) cavities have been observed at cryogenic temperatures with a DC Superconducting Quantum Interference Device (SQUID) amplifier. High $Q$ modes with bandwidths of few tens of milliHz produce thermal fluctuations with a Signal-To-Noise ratio of up to 23dB. The estimated effective temperature from the Nyquist noise is in good agreement with the physical temperature of the device, confirming the validity of the equivalent circuit model and the non-existence of any excess resonator self-noise. The measurements also confirm that the quality factor remains extremely high ($Q>10^8$ at low order overtones) for very weak (thermal) system motion at low temperatures, when compared to values measured with relatively strong external excitation. This result represents an enabling step towards operating such a high-Q acoustic device at the standard quantum limit.

\end{abstract}
\date{\today}
\maketitle

%\begin{multicols}{2}

%\section*{Introduction}

Phonon-trapping Bulk Acoustic Wave (BAW) cavity resonator technology shows great potential for use in applications that require precision control, measurement and sensing at the quantum limit\cite{Goryachev:2014aa}. This is mainly due to the relatively high mechanical frequencies and extremely high Q-factors achievable in such devices at cryogenic temperatures ($Q>10^9$), which potentially lead to extraordinarily large coherence times \cite{apl2,ScRep,quartzPRL} beyond the capability of any other competing technology compared in \cite{Kippen}. This uniqueness has been perfected for decades for precision room temperature oscillators and related devices\cite{Salzenstein:2010aa,patrice2}, culminating in $Q\times f$-products  as high as $2\cdot10^{13}$~Hz \cite{patrice}. Interestingly, it is no longer  the deficiency of the $Q$-factor,  which  halts  further  progress  in  the reduction  of  phase  fluctuations, but rather the intrinsic fluctuations of the BAW resonator itself\cite{Goryachev:2011aa,Sthal}. In particular, it has been concluded that further improvement of BAW based frequency sources could only be achieved by reducing the resonator flicker phase self-noise, since this is the dominant noise source of ultra-stable BAW oscillators both at cryogenic and room temperature\cite{Goryachev:2011aa,Goryachev:2013ly}. 

The origin of the flicker frequency noise is still poorly understood {despite its significant influence on many systems\cite{Hooge,RevModPhys.53.497} ranging from biological substances\cite{756890} to superconducting electronics\cite{PhysRevLett.103.117001,Shnirman:2007aa}.
 The influence on the frequency stability of high-performance quartz oscillators on time scales of order 1-10 seconds is well-documented\cite{Groslambert:1999dg,Sthal,560264} with several attempts to understand its origins\cite{Handel:1996aa,ShmaliyYS,Kroupa}. It is also important for other mechanical resonators such as High Overtone\cite{911733} and Thin Film\cite{4037206} BAW devices.} On the other hand, coherence times of ultra-high $Q$ quartz BAW cavities are predicted to exceed 10 seconds. Thus, the question of the influence of low Fourier frequency noise has never been raised with respect to these types of measurements due to the relatively low $Q$ factors of the mechanical resonators utilised so far\cite{cleland02,Aspelmeyer:2008qc,OConnell:2010fk}. So, this type of noise can be another limiting factor on the coherence times of these devices, as it can be for some types of superconducting qubits due to flicker noise in Josephson junctions\cite{Paladino:2014aa}. On the other hand, it has been observed that the flicker self-noise decreases with decreasing power of the incident signal, and there have been no reports of the measurement of self-noise without the carrier. Thus, one of the goals of this work was to confirm whether or not BAW devices are dominated by Nyquist noise (due to quantum or thermal fluctuations) when the carrier is not present. Thus, the observation of intrinsic Nyquist fluctuations is an important step towards preparation of a BAW resonator in the quantum ground state.
%Thus, the characterisation of the noise and fluctuations is an important milestone towards observation of quantum ground state of the BAW resonator.

The fact that the Nyquist thermal noise of BAW devices has never been directly observed experimentally is mostly due to instrumental limitations. In particular, measurements of thermal noise require low noise amplification with effective impedance matching. Whereas for low (tens-hundreds of kHz) frequencies typical for tuning-fork type devices, the goal could be achieved by utilizing high input impedance amplifiers, these type of amplifiers are non-existent at the typical cryogenic BAW frequencies (above 5-10 MHz), making such measurements rather challenging\cite{Qnoise}. Furthermore, the optomechanical approach\cite{Kippen} to thermal noise measurements can not be applied directly, since mirror coating a BVA (electrodeless) BAW resonator\cite{1537081} would immediately result in $Q$-factor degradation. With these limitations, the straightforward solution is utilisation of {a DC Superconducting Quantum Interference Device (SQUID) Amplifiers\cite{SQUIDbook,PhysRevB.86.144510} which ensure very high sensitivity. Such amplifiers have been demonstrated to operate in the quantum regime\cite{Clarke} and have been implemented in highly sensitive tests of fundamental physics \cite{PhysRevLett.104.041301,Xu:1985aa}.}
% in order to observe thermal noise current flowing through the cryogenic BAW resonator 

The measurements were made with a quartz plano-convex\cite{Tiers2} BVA\cite{1537081} SC-cut\cite{pz:1988zr} quartz BAW resonator.  The $Q$-factors of such resonators were measured to be in excess of $2$ billion at very high overtones (OT) giving $Q\times f$ products approaching $10^{18}$Hz\cite{quartzPRL} due to effective phonon trapping\cite{ScRep}. These devices exhibit resonances of the longitudinal (A-mode), fast shear (B) and slow shear (C) thickness modes, for different OT wave number $n$ and in-plane wave numbers $m$ and $p$. Thus, each resonance is referred to as X$_{n,m,p}$ where X stands for a type of vibration. Only odd overtones $n$ could be excited piezoelectrically. 

To utilise the DC SQUID Amplifier it was simply directly connected to the SQUID input coil with a few cm long twisted pair cable. 
The BAW-SQUID assembly was placed in a cryostat of a pulse-tube refrigerator. Although the pulse-tube cryocooler system is less vibrationally stable than a liquid Helium dewar, its additional shielding of the SQUID sensor, both thermal and electromagnetic\cite{Goryachev:2012jx}, proved to be sufficiently good to avoid the use of the SQUID magnetic flux control loop. 
%Both devices were placed in the thermally controlled environment at about 4K\cite{Goryachev:2012jx}. The cryogenic system is baed on a pulse-tube cryocooler, which is a less vibrationally stable environment than a liquid Helium dewar. Although cryocooler radiation and vacuum shielding provide additional screening of the SQUID from electromagnetic interference. Shielding of the SQUID sensor, both thermally and electromagneticly, proved to be sufficiently good to avoid the use of the SQUID magnetic flux control loop.
This allowed us to extend the frequency range of noise measurements by more than 3 times: from the 6 MHz limit set by the time delay in the flux lock loop to approximately 20 MHz limit set by the SQUID electronics.

The SQUID readout provides an efficient current-­to-­voltage convergence characterised by transimpedance $Z_\text{SQUID}=\frac{V_o}{I_i}$ of $1.2$MOhms with the magnetic flux noise floor of $1.09755 \mu\phi_0/\sqrt{\text{Hz}}$ at $50$kHz with $\phi_0$ being the flux quantum and the 3 dB bandwidth of approximately $2.1$MHz, which is set by the cryocable connecting room and cryogenic parts of the SQUID amplifier. The smallest detectable Root-Mean-Square (RMS) fluctuations of electric current flowing through the SQUID input coil are close to $0.5$pA$/\sqrt{\text{Hz}}$. Despite starting to lose sensitivity at 2.1MHz, the SQUID was still good enough to enable detection of the thermal noise spectra of resonances of up to $20.4$MHz (with Signal-To-Noise ratio of order $10$dB). The amplified signal is analysed with a HP89410A Vector Signal Analyser capable of providing the frequency resolution down to $1$mHz. Such resolution is required for ultrahigh-$Q$ resonance systems like cryogenic BAW resonators, which typically exhibit linewidths as small as tens of mHz\cite{ScRep,quartzPRL}. 

\begin{figure}[ht!]
     \begin{center}
            \includegraphics[width=0.45\textwidth]{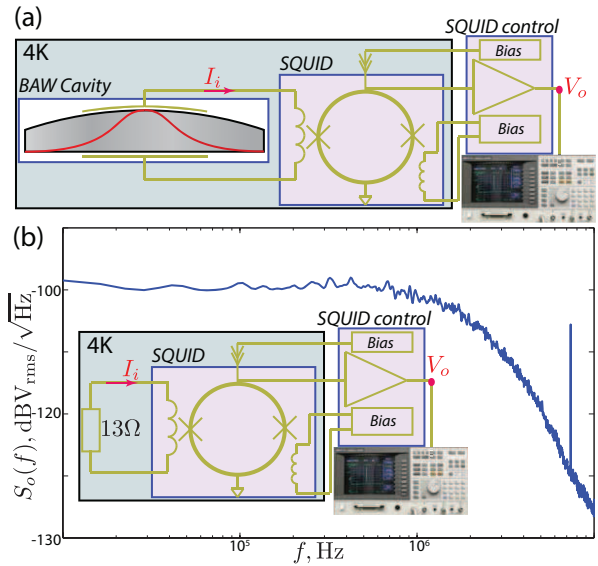}
            \end{center}
    \caption{(a) BAW thermal noise measurement setup for frequencies below 10MHz. The curvature of one of the plate faces is employed to achieve the phonon trapping with the acoustic energy distribution along the plate denoted by the red curve. (b) Resistive load measurements for system calibration.}%
   \label{setupNOISE}
\end{figure}

To ensure the frequency stability of the measurement apparatus, it was locked to a Hydrogen maser. The experimental setup is shown in Fig.~\ref{setupNOISE}, (a). For resonance frequencies above $10$MHz a down conversion technique was employed due to the frequency limitation of the vector signal analyser. In the case of higher frequencies, the noise from the output of the amplifier is down-converted by a locking amplifier with the reference frequency detuned from the BAW cavity resonance frequency by tens of Hz. The down-conversion reduces the resolution of spectral measurements, i.e. decreases the contrast of the thermal noise peak, by 3 dB. This happens due to the demodulation of both the upper and the lower spectral components surrounding the thermal noise peak to baseband. This effect can be avoided with the use of a high-speed FFT analyser.

The system has been calibrated by putting different thermal noise sources, precision resistors of different values, in the place of the quartz resonator\cite{Bara-Maillet:2014aa} (inset in Fig.~\ref{setupNOISE}, (b)). Assuming that the measured voltage noise was caused by current fluctuations in the SQUID input coil consistent with the Nyquist-Johnson theorem applied for the resistor at its surrounding temperature, one can derive the SQUID transimpedance $Z_\text{SQUID}$ relating the input coil current and the output voltage. 
%In all cases, the generated noise is consistent with the Nyquist-Johnson theorem as well as with the SQUID DC magnetic sensitivity provided by the manufacturer. 
Using the input coil coupling parameter $2\phi_0/\mu$A and the measured value of the SQUID transimpedance $Z_\text{SQUID}$, magnetic sensitivity is inferred to be $300\mu$V$/\phi_0$ that is in accordance with the manufacturer specification.
The result of these measurements is shown in Fig.~\ref{setupNOISE}, (b). The spurious signal is due to electromagnetic interference at exactly $5$MHz, which {\ is related to the distributed synchronization network. At the same time, the closest OT is   $7$~kHz away from this frequency making a significant detuning for the system with a bandwidth of $0.1$ Hz. Moreover, to ensure high frequency resolution, all measurements are done with frequency spans of about $10$~Hz. Thus, the interference of this spurious signal does not influence any of the measurements.}

\begin{figure}[ht!]
     \begin{center}
            \includegraphics[width=0.45\textwidth]{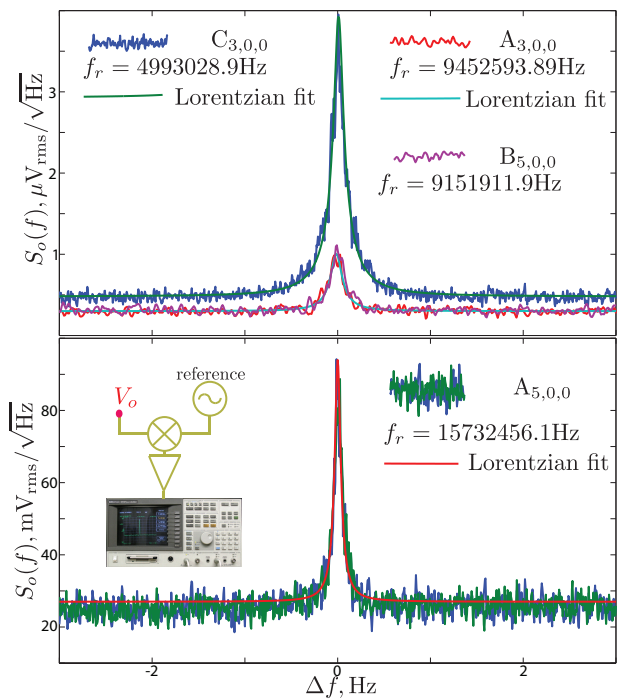}
            \end{center}
    \caption{Results of noise measurements for C$_{3,0,0}$, A$_{3,0,0}$, B$_{5,0,0}$ and A$_{5,0,0}$. The latter is measured using the downconversion as shown in inset.}%
   \label{result}
\end{figure}

It has been demonstrated that phonon trapping resonators of this size are capable of working at very high OTs with frequencies exceeding $700$MHz\cite{quartzPRL}. Although for current measurements such high frequencies could not be observed due to two reasons. Firstly, due to parasitic capacitance the photon-phonon coupling decreases with frequency. This could be overcome with the optimal electrode design, although optimal electrode size is specific to each OT. Secondly, the current SQUID bandwidth is limited to a few MHz, which only allows us to observe high-$Q$ resonances up to frequencies of order $20$MHz. In this range, 8 resonance peaks were observed. All resonance frequencies correspond to their values measured with the direct network analysis. 
Results of the measurements of noise spectra for some of the detected modes are shown in Fig.~\ref{result}.
The figure demonstrates that the shapes of the thermal noise peaks {fit well to a Lorentzian profile}, which makes it possible to infer the resonance mode bandwidths, and thus $Q$-factors {and is consistent with Nyquist thermal noise}. 

The goal of the current experiment is to estimate the mode temperatures $T_\text{mode}$ of the resonances observed. It is expected to see mode temperatures close to the physical (environment) temperature $T$. Any discrepancy implies either incorrect calibration, erroneous mathematical model or, in case of $T_\text{mode}\gg T$, additional unknown noise sources. The modes temperature estimation procedure is the following: (i) measurements of the SQUID voltage noise; (ii) estimation of the RMS current through resonator with the known SQUID impedance; (iii) calculation of the power dissipated in the resonator that is used for evaluation of mode temperature.

 For the stated purpose, we employ the commonly accepted Butterworth-Van Dyke model\cite{BVD} which represents each resonance of the BAW resonator as a series connection of inductance $L_x$, capacitance $C_x$ and resistance $R_x$. All these motional branches are connected in parallel with a shunt capacitance created by the system electrodes. Estimations in terms of this directly measurable equivalent circuit make it possible to derive the mode temperature without
making assumptions on mechanical parameters of the acoustic devices. Having in mind the fluctuation-dissipation theorem, we postulate that the motional resistance responsible for the dissipation in the equivalent circuit is also a source of random fluctuations. Thus, each resonance branch is associated with its own noise source. At thermal equilibrium, the variance of this noise is proportional to the system ambient temperature, and the spectral density of noise power in Watts/Hz, $S_N$, is given by the Planck's equation applied to a resistor at physical temperature $T$\cite{IEEEnoise}:
\begin{equation}
	\label{R001SW}
		S_N=k_BT\frac{\beta\hbar\omega}{e^{\beta\hbar\omega}-1},
\end{equation}
where $\beta=1/(k_BT)$, $k_B$ is the Boltzmann constant. This  expression is widely considered to be the exact formula, to which the Rayleigh-Jeans form is an approximation 
valid for small $\beta\hbar\omega$\cite{IEEEnoise}. Since the discrepancy between the two laws can only account for up to 5\% for the frequency range of investigation, our calculations are based on the approximate formula. %Thus, the power spectral density (PSD) of thermal fluctuations of current in the input coil of the SQUID amplifier is predicted to be,
%\begin{equation}
%	\label{R002SW}
%		S_i(\omega) =\Big(\frac{4k_BT_\text{mode}}{R_x}\Big)\Big(\frac{1}{e^{\beta\hbar\omega}-1}\Big)\Big(\frac{1}{1+(\omega\tau)^2}\Big),
%\end{equation}
%where the last fraction accounts for the frequency response of the SQUID input circuit and $\tau$ is the corresponding time constant. The PSD of the voltage noise at the SQUID output is then given by $S_o(\omega)=Z_\text{SQUID}^2S_i(\omega)$. 

The power spectral densities (PSD) of the voltage noise at the SQUID output is related to the PSD of the current fluctuations in the input coil through the transimpedance: $S_o(\omega)=Z_\text{SQUID}^2S_i(\omega)$.
Based on this relation, the power of the current fluctuations in the SQUID input coil, $p$, induced by a given mode (assuming Lorentzian lineshape), can be calculated from the experimental data using the following relation:
\begin{equation}
	\label{R003SW}
		\displaystyle p =\int\limits_{0}^{\infty}R_xS_i(f)df =\frac{\omega_r}{4Q}R_x\frac{V_p^2}{Z^2_\text{SQUID}},%
\end{equation}
where $\omega_r=2\pi f_r$, $Q$ and $R_x$ are the mode angular frequency, quality factor and motional resistance, while $V_p$ is the maximum RMS value of voltage (in V/$\sqrt{\text{Hz}}$) at the SQUID output at the resonance frequency. The values of $\omega_r$ and $Q$ could be measured either by fitting the curves of the measured noise spectra (Fig.~\ref{result}) or by impedance analysis measurements with an excitation signal\cite{apl1,apl2,ScRep}. Whereas the angular frequency estimation by both methods always coincides with a high degree of precision, the quality factor measurements show some discrepancy. In either case, the motional resistance can be measured only by the impedance analysis technique, which requires an extra stage in the procedure of calibration.
The resonance parameters, which are required to estimate effective noise temperature for each mode are given in Table~\ref{modesT}. The results of the experimental estimations of the power of current fluctuations in the SQUID input coil given by Eq.~(\ref{R003SW}) are used to calculate the noise energy stored in the resonator:
\begin{equation}
	\label{R004SW}
		E = \tau_\text{mode}pH_{\text{inp}}^{-1}= R_x\frac{V_p^2}{4Z^2_\text{SQUID}}\Big({1+(\omega\tau)^2}\Big),
\end{equation}
where $H_{\text{inp}}$ is the transfer function that accounts for the frequency response of the SQUID input circuit with $\tau$ being the corresponding time constant, $\tau_\text{mode}=\frac{Q}{\omega_r}$ is the mode relaxation time.
The experimental estimation made with this relation is then compared to the theoretical predictions of $S_N$ given by Eq.~(\ref{R001SW}) to derive the mode temperature $T=T_\text{mode}$ as a parameter.

In addition to the measurements at $4$K, resonator thermal noise measurements were also made at $50$mK. For these measurements the BAW resonator was placed at the mK-stage of a dilution refrigerator whereas the SQUID amplifier is put on a 4K plate of the same system. The devices were connected via a coaxial cable. 

\begin{table}[t]
\caption{Parameters of an X$_{m,n,p}$ resonance at 4K: $f_r$ - resonance frequency, $Q_\text{fit}$ - quality factor from the Lorentz fit, $V_p$ - noise peak value, $R_x$ and $Q_Z$ are motional resistance and quality factor as measured by the impedance analyser.}
\centering
\begin{tabularx}{\columnwidth}{XXXXXX}
\hline
\hline
X$_{n,m,p}$ & $f_r,$ MHz & $Q_\text{fit},10^{7}$  & $V_p,\frac{\mu\text{V}}{\sqrt{\text{Hz}}}$  & $R_x, \Omega$ &$Q_{Z},10^{7}$ \\
   %  &  X$_{nmp}$ &  &  & \\
\hline
 C$_{3,0,0}$ & $4.993$& $4.9$  & $4$ & $3$   &  $4.4$ \\
 C$_{5,0,0}$ & $8.392$&   &  &  $5.36$ & $10.7$  \\
 B$_{3,0,0}$ & $5.505$&   &  & $2.9$   & $4.84$  \\
 B$_{5,0,0}$ & $9.152$& $10$  & $1.1$  & $5.9$   &  $6.4$ \\
 B$_{5,m,p}$ & $9.247$& $10$  &  &  $7.9$  & $11.9$  \\
 B$_{5,m,p}$ & $9.287$& $10$  & $0.87$  &  $5.92$  & $25.8$  \\
 A$_{3,0,0}$ & $9.452$& $9.4$  & $1$  & $6$   &  $6$ \\
A$_{5,0,0}$ & $15.732$& $26$  &  & $2.3$   &  $30$ \\
\hline
 C$_{3,0,0}$ & $4.993$& $2.9$  & $0.74$ & $2.3$   &  $3$ \\
  $@$50mK &&    &    &   \\
\hline
\end{tabularx}
\label{modesT}
\end{table}

Difference between estimation of $Q$ with two different techniques (direct impedance measurements and Lorentzian fit to the thermal noise spectra) is mainly due to the weakness of the thermal noise peak. In spite of some discrepancy both methods demonstrates the same order of magnitude. This fact confirms that the $Q$ values of BAW cavities measured with the relatively strong excitation signal ($\sim 10^{10}$ acoustic phonons)\cite{ScRep, quartzPRL} hold the same at very weak excitation level, even in the absence of any external signal ($<10^4$ thermal phonons). 

\begin{figure}[ht!]
     \begin{center}
            \includegraphics[width=0.45\textwidth]{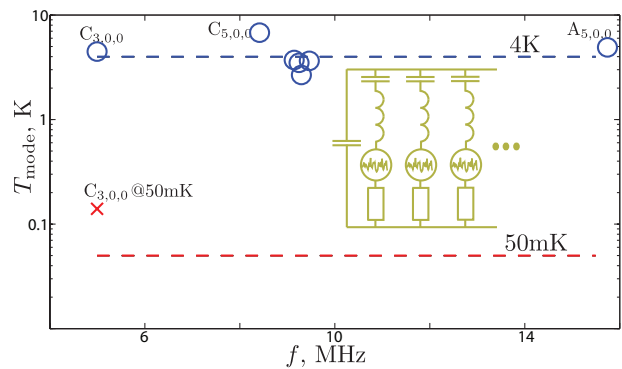}
            \end{center}
    \caption{Estimations of the mode temperatures compared to the ambient values. The inset shows the equivalent circuit model.}%
   \label{temper}
\end{figure}

Fig.~\ref{temper} demonstrates the mode effective temperature as a function of resonance frequency. Whereas measurements made at 4K show mode temperature to be close to the ambient, mK results depict significant discrepancy between the two values. Only for the C$_{3,0,0}$ mode shown was relatively close to the ambient temperature of the quartz crystal. The results obtained for the other higher frequency modes were inconclusive due to poor signal to noise ratio. The mK experiment is susceptible to additional noise induced from the cables connecting the BAW resonator at 50 mK and the SQUID amplifier at 4K. There are no circulators at these frequencies to isolate the two devices, so it is likely that thermal noise at 4 K is causing back action on the BAW cavity. The next step will be to cool the SQUID to mK along with the sample to circumvent this problem.

 %Table~\ref{modesT} presents measured parameters of these resonances. %, the following resonances have been measured: A$_{3,0,0}$, B$_{3,0,0}$, C$_{3,0,0}$, A$_{5,0,0}$, B$_{5,0,0}$, C$_{5,0,0}$, B$_{5,0,2}$ and B$_{5,2,0}$.

%The results for the C$_{3,0,0}$ mode at this temperature are shown in Table~\ref{modesT}.

In summary, thermal noise fluctuations of several low order thickness modes of a high-$Q$ phonon-trapping acoustic cavity has been measured with a low-noise SQUID amplifier. For each mode, effective noise temperature has been estimated. The results at $4$K are in reasonable agreement with the device physical temperature. This fact confirms validity of both the Butterworth-Van Dyke equivalent circuit for description of thermal fluctuations in the piezoelectric acoustic resonator at cryogenic temperatures. The results also confirm very high $Q$ values of acoustic modes at extremely low powers. %The technique could be improved by increasing the amplifier bandwidth. 
The use of the new generation of DC SQUID systems cooled to mK temperatures with the bandwidth in excess of 100 MHz, along with the cross-correlation signal processing, would enable us to access and investigate the noise properties of the high-order overtone acoustic resonances at the quantum limit. Such resonances were shown to posses the $Q$-factors in excess of one billion which makes them ideal candidates for observation of quantum behavior of macroscopic objects.

%S_o(f), \text{dBV}_\text{rms}/\sqrt{\text{Hz}}

This work was supported by the Australian Research Council Grant No. CE11E0082 and FL0992016. Authors are thankful to Jean-Pierre Aubry for the resonator samples. 
\hspace{10pt}

%\bibliography{biblioBAW}

\end{document}